\documentclass[twocolumn,prl,superscriptaddress,preprintnumbers,amsmath,amssymb]{revtex4}

\newcommand{\m}[1]
{\mathrm{#1}}

\usepackage{graphicx}
\usepackage{dcolumn}
\usepackage{bm}
\usepackage{times}
\usepackage{color}

\begin{document}

\bibliographystyle{unsrt}
\preprint{M. Kroner {\em et al.} \today}

\title{Optical detection of single electron spin resonance in a quantum dot}

\author{Martin Kroner}
\affiliation{Center for NanoScience and Fakult\"{a}t f\"{u}r Physik, Ludwig-Maximilians-Universit\"{a}t, 80539 M\"{u}nchen, Germany}

\author{Kathrina M. Weiss}
\affiliation{Center for NanoScience and Fakult\"{a}t f\"{u}r Physik, Ludwig-Maximilians-Universit\"{a}t, 80539 M\"{u}nchen, Germany}

\author{Benjamin Biedermann}
\affiliation{Center for NanoScience and Fakult\"{a}t f\"{u}r Physik, Ludwig-Maximilians-Universit\"{a}t, 80539 M\"{u}nchen, Germany}

\author{Stefan Seidl}
\affiliation{Center for NanoScience and Fakult\"{a}t f\"{u}r Physik, Ludwig-Maximilians-Universit\"{a}t, 80539 M\"{u}nchen, Germany}

\author{Stefan Manus}
\affiliation{Center for NanoScience and Fakult\"{a}t f\"{u}r Physik, Ludwig-Maximilians-Universit\"{a}t, 80539 M\"{u}nchen, Germany}

\author{Alexander Holleitner}
\affiliation{Center for NanoScience and Fakult\"{a}t f\"{u}r Physik, Ludwig-Maximilians-Universit\"{a}t, 80539 M\"{u}nchen, Germany}

\author{Antonio Badolato}
\affiliation{Materials Department, University of California, Santa
Barbara, California 93106, USA}

\author{Pierre M. Petroff}
\affiliation{Materials Department, University of California, Santa
Barbara, California 93106, USA}

\author{Brian D. Gerardot}
\affiliation{School of Engineering and Physical Sciences, Heriot-Watt 
University, Edinburgh EH14 4AS, UK}

\author{Richard J. Warburton}
\affiliation{School of Engineering and Physical Sciences, Heriot-Watt 
University, Edinburgh EH14 4AS, UK}

\author{Khaled Karrai}
\affiliation{Center for NanoScience and Department f\"{u}r Physik, Ludwig-Maximilians-Universit\"{a}t, 80539 M\"{u}nchen, Germany}

\date{\today}



\begin{abstract}
We demonstrate optically detected spin resonance of a single electron confined to a self-assembled quantum dot. The dot is rendered dark by resonant optical pumping of the spin with a coherent laser. Contrast is restored by applying a radio frequency (rf) magnetic field at the spin resonance. The scheme is sensitive even to rf fields of just a few $\mu$T. In one case, the spin resonance behaves exactly as a driven 3-level quantum system (a $\lambda$-system) with weak damping. In another, the dot exhibits remarkably strong (67\% signal recovery) and narrow (0.34 MHz) spin resonances with fluctuating resonant positions, evidence of unusual dynamic processes of non-Markovian character.  
\end{abstract}

\maketitle

The quantum mechanical control of few-level systems is a major challenge for the development of novel computation and communication schemes based on quantum states. Solid state-based nanostructures can be tailored and even tuned in situ, significant advantages over conventional quantum systems such as atoms and ions. Furthermore, in a strongly quantized solid state system, electron spin is remarkably robust as the quantization suppresses the phonon-related spin relaxation \cite{Golovach,Krautvar,Elzerman}, adding weight to proposals using spin as a qubit \cite{Loss}. Recently, spin relaxation times as long as $T_{1}\sim 1$ s \cite{Amasha} and a lower bound on the coherence time $T_{2}$ of 1 $\mu$s \cite{Petta} have been established on electrostatically-defined quantum dots. It is clearly of fundamental importance to develop spin resonance schemes with single spin resolution. In the longer term, spin resonance provides the capability of performing arbitrary spin rotations in the Bloch sphere with high fidelity; in the shorter term, it provides unique insights into the complex spin interactions in the solid state environment. Single spin resonance has been achieved on an electrostatically-defined  quantum dot with electrical detection and has led to the observation of Rabi flopping, equivalently single spin rotations \cite{Koppens}. An alternative is to detect the spin resonance optically. This is potentially very sensitive because of the in-built amplification of $\sim 10^{5}-10^{7}$: absorption of a microwave photon leads to absorption of an optical photon. While optically detected single spin resonance on the NV$^{-}$ center, a deep impurity level in diamond, is established \cite{Jelezko}, optically detected single spin resonance on an extended system, a quantum dot, has not been demonstrated before. The crucial difference is the length scale: we report here a spin resonance of a single electron with a wave function extending over $\sim 10^{5}$ atoms.

\begin{figure}
\begin{center}
\includegraphics[angle=0,width=\columnwidth]{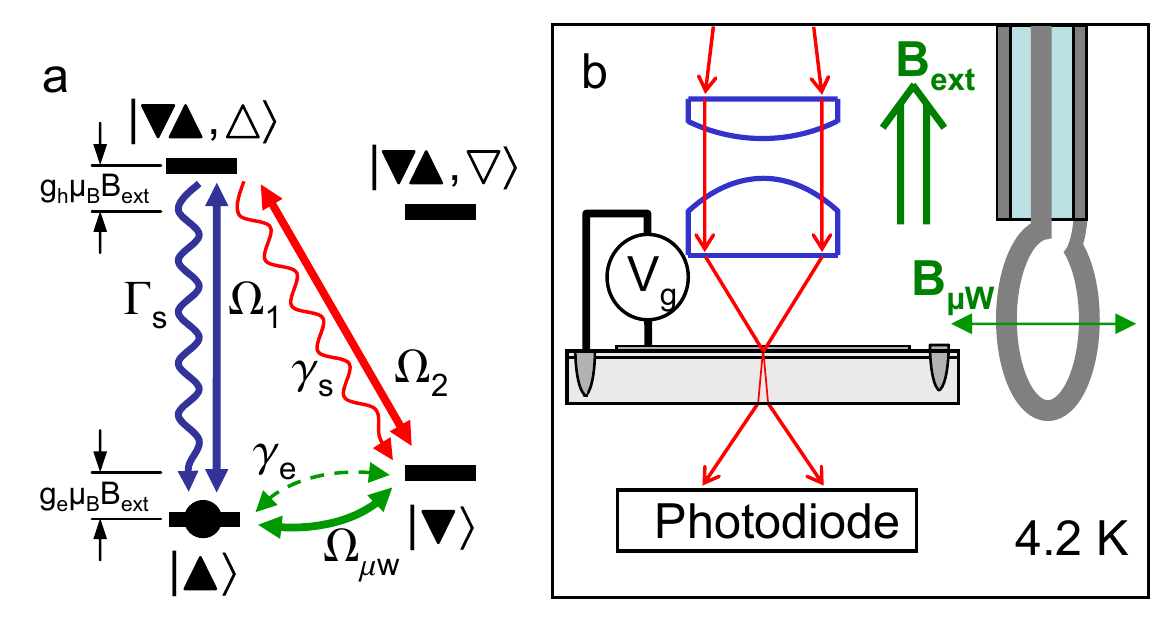}
\end{center}
\caption{\label{scheme}(a) Level scheme for optically detected spin resonance. The electron spin states, $|\uparrow\rangle$ and $|\downarrow\rangle$, are split by the electron Zeeman energy $g_\m{e} \mu_\m{B}B_\m{ext}$ ($g_\m{e}<0$). $|\uparrow\downarrow,\Uparrow\rangle$ denotes the spin-up exciton state, X$^{1-}$. The $\sigma^{+}$-polarized transition $|\uparrow\rangle\leftrightarrow|\uparrow\downarrow,\Uparrow\rangle$ is driven on resonance with a coherent laser. The spin resonance transition, $|\downarrow\rangle \leftrightarrow |\uparrow\rangle$, is driven with an oscillating magnetic field $B_\m{\mu W}$ at right angles to the dc external magnetic field, $B_\m{ext}$. Incoherent processes are spontaneous radiative decay, $|\uparrow\downarrow,\Uparrow\rangle\rightarrow |\uparrow\rangle$ (fast), $|\uparrow\downarrow,\Uparrow\rangle\rightarrow |\downarrow\rangle$ (slow); and spin relaxation, $|\downarrow\rangle\leftrightarrow|\uparrow\rangle$ (very slow). (b) Schematic of experimental setup. The laser excitation at wavelengths around 950 nm is focused onto the sample with an objective with numerical aperture 0.5 and gives a spot size of $\sim 1$ $\mu$m at the sample. A gate voltage $V_\m{g}$ is applied between surface Schottky gate and the back contact and controls the quantum dot charge via Coulomb blockade and within one charge plateau, the energy of the exciton transition via the Stark effect. The transmitted light is detected with an in situ photodiode. A dc magnetic field, $B_\m{ext}$, is applied perpendicular to the sample. An ac magnetic field, $B_\m{\mu W}$, is generated with a closed loop antenna of diameter 2 mm positioned 2 mm along its axis from the quantum dot. The loop is connected to a microwave oscillator via a semi-rigid high frequency cable. The objective, sample and antenna are all at 4.2 K.} 
\end{figure}

Our electron spin resonance (ESR) scheme is shown in Fig.\ \ref{scheme}(a). A single electron is confined to the dot and a magnetic field is applied, splitting the electron spin states, $|\downarrow\rangle$ and $|\uparrow\rangle$, by the Zeeman energy. The first step is to project the electron into, say, the $|\downarrow\rangle$ state. We do this with optical pumping: a laser drives the $|\uparrow\rangle \leftrightarrow |\uparrow\downarrow,\Uparrow\rangle$ transition where $|\uparrow\downarrow,\Uparrow\rangle$ represents the X$^{1-}$ exciton consisting of two spin-paired electrons and a spin-up hole. Spontaneous emission damps the laser-driven oscillations and, in the presence of some symmetry breaking, can project the electron into the $|\downarrow\rangle$ state where the population is shelved. This method of spin initialization is highly efficient provided the spin relaxation rate from $|\downarrow\rangle$ to $|\uparrow\rangle$ is slow. Its signature is the disappearance of a signal related to Rayleigh scattering of the optical laser \cite{Mete,Martin,Jan}. The second step is to apply a radio frequency (rf) magnetic field at the Zeeman frequency. The rf field drives the spin resonance transition, $|\downarrow\rangle \leftrightarrow |\uparrow\rangle$. This causes the $|\uparrow\rangle$ state to be re-populated, re-establishing the Rayleigh scattering signal. The scheme is a contemporary application of magnetic resonance developed originally with Hg atoms \cite{Brossel}, replacing a huge ensemble of atoms with a single quantum dot, an incoherent source with a laser and fluorescence detection with Rayleigh scattering.

We perform spin resonance on single InAs/GaAs self-assembled quantum dots. The dots are embedded in a field effect structure which allows for controlled charging with single electrons via a tunneling interaction with an electron reservoir \cite{Richard,Stefan}. In the present device, the tunneling barrier is 25 nm thick and the back contact is a two-dimensional electron gas with carrier concentration $10^{12}$ cm$^{-2}$. The sample is mounted in a 4 K optical microscope, a magnetic field of $B_\m{ext}=0.5$ T is applied in the growth direction (Faraday geometry), and a voltage $V_\m{g}$ is applied to a surface Schottky barrier in order to trap a single electron in a particular dot, Fig.\ \ref{scheme}(b). The interaction with a narrowband laser tuned to the fundamental cross-gap transition is detected in transmission: a dip in transmission with linewidth $\sim 2$ $\mu$eV is observed on resonance \cite{Alex}. The quantum dot is tuned relative to the constant laser energy by sweeping $V_\m{g}$ which shifts the exciton through the Stark effect. The microwave field is generated by a single loop antenna with a geometry designed to emit over a broad frequency spectrum, Fig.\ \ref{scheme}(b). The investigated quantum dot lies close to the antenna on its symmetry axis ensuring that the dot experiences only the near-field of the antenna, which contains a magnetic but no electric field component. From the geometry and the electrical characteristics of the setup, we estimate the ac magnetic field $B_\m{\mu W}$ to be a few $\mu$T. 

\begin{figure}
\begin{center}
\includegraphics[angle=0,width=\columnwidth]{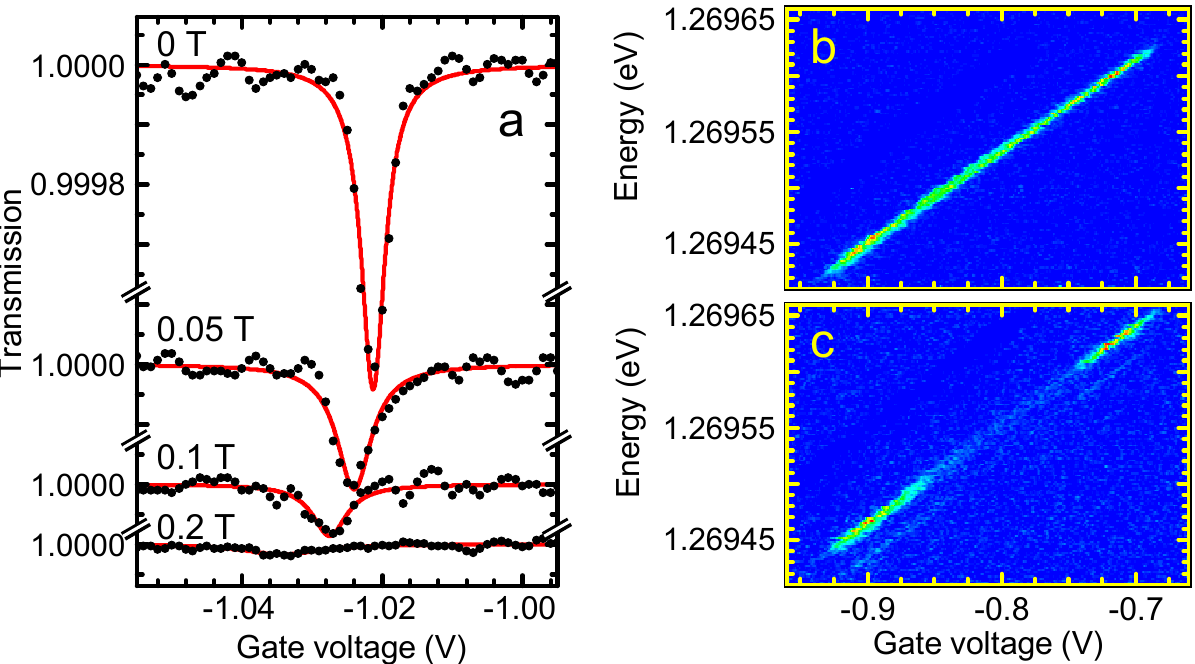}
\end{center}
\caption{\label{spinpumping}Spin shelving via optical pumping. (a) shows differential transmission data on a single quantum dot at 4.2 K as a function of the applied magnetic field, $B_\m{ext}$, recorded with a $V_\m{g}$ at the center of the charging plateau. The contrast disappears as $B_\m{ext}$ increases. (b), (c) color scale plots of the $V_\m{g}$-dependence. (b) At $B_\m{ext}=0$, the optical signal is maintained across the plateau; (c) at $B_\m{ext}=0.5$ T, the optical signal is suppressed in the plateau center signifying spin shelving, but recovers at the plateau edges signifying rapid spin relaxation via cotunneling.}
\end{figure}

The primary signature of optical spin pumping is a loss of transmission signal as a magnetic field is applied \cite{Mete,Martin,Jan}, Fig.\ \ref{spinpumping}. The hyperfine interaction, in particular the effect of the Overhauser field on the electron spin, plays a dual role here \cite{Jan}. First, at low magnetic fields, the fluctuating nuclear field induces rapid electron spin relaxation \cite{Jan,Merkulov}. As the field increases, this relaxation mechanism is suppressed on account of the energetic mismatch in nuclear and electronic Zeeman energies. Secondly, the in-plane Overhauser field is responsible for the symmetry breaking required for the ``forbidden" $|\uparrow\downarrow,\Uparrow\rangle \rightarrow |\downarrow\rangle$ transition, Fig.\ \ref{scheme}(a). The in-plane field admixes the $|\downarrow\rangle$ and the $|\uparrow\rangle$ states; equivalently it tilts the electron (but not the exciton) quantization axis away from the $z$-axis. The in-plane Overhauser field is $\sim 30$ mT for these dots \cite{Jan} such that the spin admixture is small at $B_{ext}=0.5$ T but nevertheless sufficient for spin pumping. The spin relaxation rate can also be controlled via $V_\m{g}$: at the edges of the Coulomb blockade plateau, spin relaxation is rapid via a spin-swap with the back contact, a co-tunneling process \cite{Mete,Martin,Jan,Jason}. This inhibits spin pumping and causes the transmission signal to reappear at the plateau edges, Fig.\ \ref{spinpumping}. 

\begin{figure}
\begin{center}
\includegraphics[angle=0,width=\columnwidth]{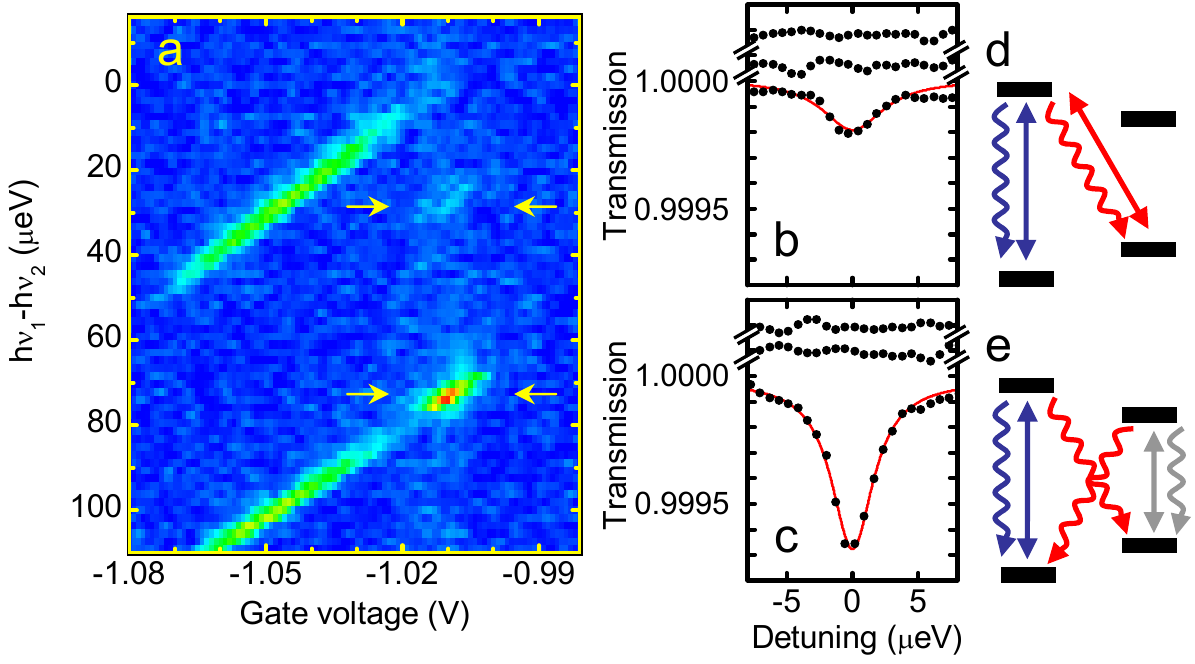}
\end{center}
\caption{\label{2colour}Optical characterization with 2 coherent lasers. The energy of laser 1 is chosen to come into resonance with the higher energy Zeeman transition, $|\uparrow\rangle \leftrightarrow|\uparrow\downarrow,\Uparrow\rangle$, at $V_\m{g}=-1.01$ V where the transmission contrast is immeasurably small owing to spin shelving in the $|\downarrow\rangle$ state. The power of laser 1 is 1 nW. The energy of a second laser, $h\nu_{2}$, with power 1 nW is red-shifted relative to laser 1 and the differential transmission is recorded as a function of $V_\m{g}$. Both lasers are linearly polarized. The process is repeated for different $h\nu_{2}$. (a) shows the data as a color plot. (b) and (c) show the transmission spectra for laser 1 alone, laser 2 alone and for laser 1 and laser 2 (offset for clarity) for the two resonances marked in (a). (d) and (e) are the respective interpretation of the resonances in terms of the level diagram, Fig.\ \ref{scheme}.} 
\end{figure}

One of the challenges in the ESR experiment is that the electron g-factor, $g_\m{e}$, for a given quantum dot is unknown. $g_\m{e}$ is strongly dependent on the detailed morphology of the quantum dot \cite{Gio}. Furthermore, conventional laser spectroscopy measures only the sum of the electron and hole Zeeman energies, and the hole Zeeman energy is typically two or three times larger than the electron Zeeman energy. Compounding this, the ESR is potentially very narrow in frequency space given the highly coherent spin at $B_\m{ext}\sim 0.5$ T. This represents a spectral ``needle in a haystack" problem. To solve it, we have developed a laser spectroscopy technique to determine $g_\m{e}$. The concept, Fig.\ \ref{2colour}(d), is to apply two laser fields, the first on resonance with the strong $|\uparrow\rangle\leftrightarrow|\uparrow\downarrow,\Uparrow\rangle$ transition which projects the spin into the $|\downarrow\rangle$ state, and the second tuned in energy to the weak $|\downarrow\rangle \leftrightarrow|\uparrow\downarrow,\Uparrow\rangle$ transition. On resonance, the second laser frustrates the spin shelving induced by the first laser, leading to a recovery of the optical transmission signal. In practice, we choose the frequency of the first laser $\nu_{1}$ such that it comes into resonance with the higher energy Zeeman transition, $|\uparrow\rangle\leftrightarrow|\uparrow\downarrow,\Uparrow\rangle$, at a $V_\m{g}$ far from the co-tunneling regime ($-1.01$ V in Fig.\ \ref{2colour}) and the transmission signal is therefore quenched by efficient spin pumping. The frequency of the second laser, $\nu_{2}$, is then gradually red-shifted relative to the first. Both lasers are linearly-polarized so that all possible circularly-polarized transitions can be pumped and both lasers are incident on the same transmission detector. For each $\nu_{2}$ we scan the gate voltage, Fig.\ \ref{2colour}(a). The two regions of high transmission contrast between $-1.06$ and $-1.03$ V represent the interaction of the second laser with the two Zeeman-split transitions, $|\uparrow\rangle \leftrightarrow|\uparrow\downarrow,\Uparrow\rangle$ and $|\downarrow\rangle \leftrightarrow|\uparrow\downarrow,\Downarrow\rangle$, in the co-tunneling regime where the contrast is large. For $V_\m{g}\ge -1.03$ V, spin shelving starts and the contrast from the strong Zeeman transitions quenches. However, there are two values of $h\nu_{2}$ where contrast is re-established at $V_\m{g}=-1.01$ V signifying double resonances, arrows in Fig.\ \ref{2colour}(a). Fig.\ \ref{2colour}(b),(c) show line cuts through the two resonances. The spectra were measured with each laser separately (no visible transmission dip) and then with both together (transmission dip) and are interpreted with the level diagrams of Fig.\ \ref{2colour}(d),(e). In (c) and (e), laser 1 pumps the higher energy Zeeman transition, and laser 2 pumps the lower energy Zeeman transition. Laser 2 frustrates the spin shelving from laser 1 and vice versa, leading to a recovery in contrast, a repumping phenomenon \cite{Mete}. In (b) and (d) however, laser 2 pumps the weaker cross-transition, and at this resonance, $h\nu_{1}-h\nu_{2}$ corresponds to the electron Zeeman energy allowing its experimental determination. The power of this scheme is that, by monitoring the strong $|\uparrow\rangle \leftrightarrow|\uparrow\downarrow,\Uparrow\rangle$ transition in the spin pumping regime, we can detect the presence of the weak $|\downarrow\rangle\leftrightarrow|\uparrow\downarrow,\Uparrow\rangle$ transition which is completely hidden either in conventional laser spectroscopy or photoluminescence characterization. In fact the detection of the ESR proceeds in a similar way, in this case the spin resonance frustrates the spin shelving.

\begin{figure}
\begin{center}
\includegraphics[angle=0,width=\columnwidth]{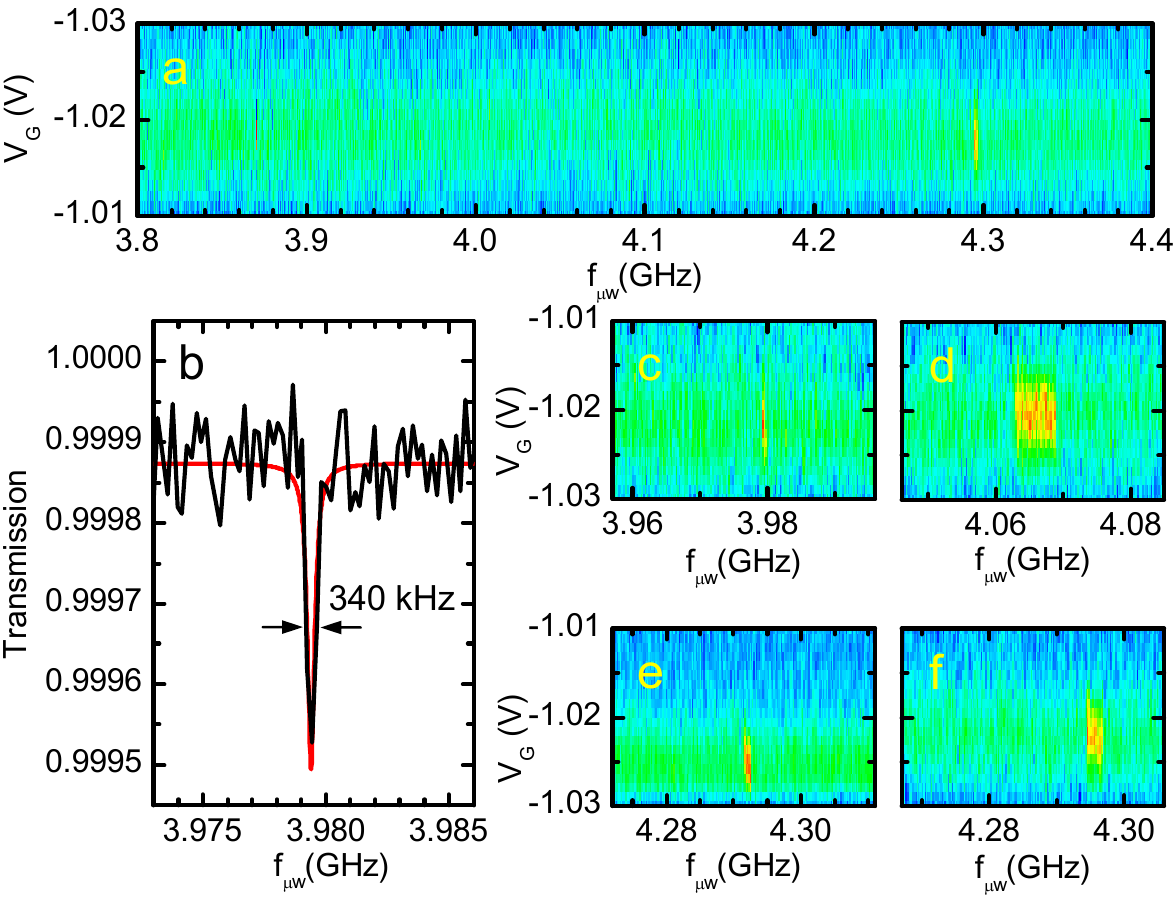}
\end{center}
\caption{\label{esrdotA} Optically detected spin resonance for dot A at $B_\m{ext}=0.5$ T. Color scale plot of the optical transmission signal with microwave frequency, $f_\m{\mu W}$, along the x-axis, gate voltage (equivalently optical detuning, $\delta$) along the y-axis. The Stark shift $d\delta/d V_\m{g}$ is 0.9 $\mu$eV/mV. For each microwave frequency, the gate voltage is swept from $-1.01$ to $-1.03$ V. A weak optical resonance can be made out at all microwave frequencies proving that the dot remains in resonance with the optical laser. The strong signal close to 4.3 GHz is the electron spin resonance (ESR). (c)-(f) other experimental runs on the same dot under identical conditions; (b) optical transmission versus microwave frequency at zero optical detuning from (c) showing ESR close to 3.98 GHz.}
\end{figure}

\begin{figure}
\begin{center}
\includegraphics[angle=0]{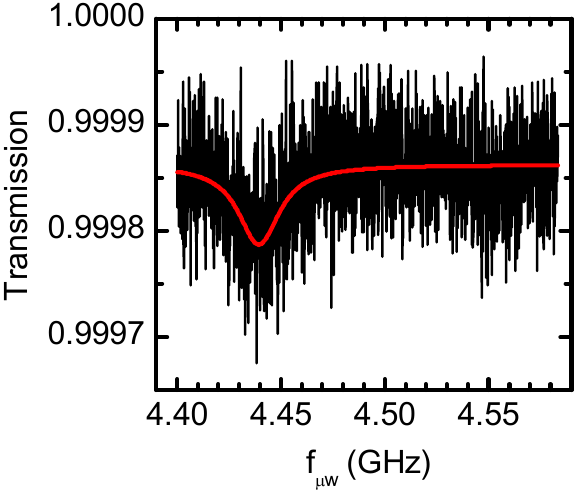}
\end{center}
\caption{\label{esrdotB} Optically detected spin resonance for dot B at 0.5 T. Optical transmission signal at zero optical detuning is plotted against microwave frequency, $f_\m{\mu W}$. The black line corresponds to the experimental data; the red line the theory. The theory uses parameters $\hbar \Gamma_\m{s}=1.0$ $\mu$eV (0.66 ns radiative decay time), $\gamma_\m{s}=0.0008 \Gamma_\m{s}$, $\hbar\gamma_\m{e}=2.4$ peV ($T_{1}=0.27$ ms), $\Omega_{1}=0.30$ $\mu$eV, $\Omega_\m{\mu W}=0.36$ neV ($B_\m{\mu W}=11$ $\mu$T, $g_\m{e}=-0.56$).}
\end{figure}

From the laser spectroscopy, we calculate the electron g-factor of the dot in Fig.\ \ref{2colour} to be $-0.56 \pm 0.05$. The prediction is that at $B_\m{ext}=0.5$ T, the ESR frequency is $3.9\pm 0.3$ GHz. To search for the ESR, we scanned the rf over a 1 GHz bandwidth with a resolution of 0.1 MHz. This is a time-consuming task. To ensure that the optical laser remains in resonance with the $|\uparrow\rangle\leftrightarrow|\uparrow\downarrow,\Uparrow\rangle$ transition throughout, we decided not to operate in the center of the plateau where the transmission signal is buried in the noise, but in a regime of $V_\m{g}$ close to the plateau edges where co-tunneling is sufficiently strong to give us a small but measurable optical signal. For each rf, we swept the quantum dot through the optical laser by sweeping $V_\m{g}$. Fig.\ \ref{esrdotA} is a contour plot of $V_\m{g}$ (optical detuning) versus rf for dot A; Fig.\ \ref{esrdotB} optical signal versus rf at zero optical detuning for dot B. For all rf, the optical resonance can be just made out in Fig.\ \ref{esrdotA}, demonstrating that the optical resonance is maintained throughout. However, at very specific microwave frequencies, there is an increase in optical signal corresponding to ESR. For dot A, the signal recovers to 67\% of its value at $B_\m{ext}=0$ and the ESR linewidth is extremely small, 0.34 MHz. For dot B, the signal recovery is smaller, 12\%, and the ESR linewidth is larger, 24 MHz. 

The scheme in Fig.\ \ref{scheme}(a) can be treated quantum mechanically with the standard techniques of quantum optics. We consider 3 levels, $|\uparrow\rangle$, $|\downarrow\rangle$ and $|\uparrow\downarrow,\Uparrow\rangle$ with a coherent optical coupling between $|\uparrow\rangle$ and $|\uparrow\downarrow,\Uparrow\rangle$ with Rabi energy $\hbar \Omega_{1}$ and either a coherent optical coupling between $|\downarrow\rangle$ and $|\uparrow\downarrow,\Uparrow\rangle$ (two-color experiment) or a coherent rf coupling between $|\uparrow\rangle$ and $|\downarrow\rangle$ (ESR) with Rabi energy $\hbar\Omega_\m{\mu W}=g_\m{e}\mu_\m{B}B_\m{\mu W}$ where $\mu_\m{B}$ is the Bohr magneton. The dynamics within this 3-level system are described, including decay terms, with a master equation for the density matrix including decay terms: spontaneous radiative recombination from $|\uparrow\downarrow,\Uparrow\rangle$ to $|\uparrow\rangle$ at rate $\Gamma_\m{s}$; spontaneous radiative recombination from $|\uparrow\downarrow,\Uparrow\rangle$ to $|\downarrow\rangle$ at rate $\gamma_\m{s}$ ($\gamma_\m{s}\ll \Gamma_\m{s}$); and spin relaxation $|\uparrow\rangle \leftrightarrow |\downarrow\rangle$ at rate $\gamma_\m{e}$. All these processes are assumed to be Markovian in character. The integration time of the experiment, $\sim 1$ s per data point, is assumed to be the longest time in the problem, leading to a description of the experiment with the steady state limit. The transmission signal depends on the optical susceptibility which depends on an optical coherence, an off-diagonal component of the density matrix. $\Gamma_\m{s}$ is known from the measured radiative decay rates on these dots and $\hbar\Omega_{1}$ is known from saturation curves of the optical resonance without spin pumping. We determine $\gamma_\m{s}/\Gamma_\m{s}=(\sqrt{2}B_\m{N,x}/2B_\m{ext})^2=0.08$\% from a root-mean-square Overhauser field in the $x$-direction of $B_\m{N,x}=20$ mT \cite{Jan}. To fit the ESR, Fig.\ \ref{esrdotB}, we require $\hbar\gamma_\m{e}=2.4$ peV ($T_{1}=0.27$ ms, limited by co-tunneling) and $\hbar\Omega_\m{\mu W}=0.36$ neV corresponding to $B_\m{\mu W}=11$ $\mu$T for $g_\m{e}=-0.56$. This $B_\m{\mu W}$ agrees reasonably with the rough estimate based on Fig.\ \ref{scheme}. Fig.\ \ref{esrdotB} shows that we achieve an excellent fit to the ESR data for dot B. For this dot, the optically detected ESR signal is just 12\% but we note that the theory predicts that the optical signal will increase to close to 100\% on increasing $B_\m{\mu W}$ by an order of magnitude. This is clearly possible with a microscopic rather than a macroscopic antenna \cite{Koppens}.

In the light of these calculations, the large signals and narrow linewidths of dot A, Fig.\ \ref{esrdotA}, are fascinating. It is impossible to reproduce these results with the 3-level calculation without reducing the optical power by an unrealistic factor of $\sim 100$. An assumption in the model must be broken for this dot. The experiment itself provides a clear pointer: repeats of the measurement on dot A show that the ESR jumps in frequency from one run to the next but yet maintains its narrow linewidth, Fig.\ \ref{esrdotA}(a),(c). In other cases, Fig.\ \ref{esrdotA}(d),(e),(f), the ESR has a strange lineshape, with hints that the ESR locks on to the rf driving field over a band of frequencies. In complete contrast, ESR traces for dot B are reproducible from one scan to the next. The results therefore point to a breakdown in the Markovian approximation for dot A but not for dot B. The difference in the ESR of the two dots is not understood and cannot be anticipated from the very similar optical characteristics. The most obvious culprit for dot A is the nuclear magnetic field: although the hyperfine interaction is strongly suppressed at $0.5$ T, it is not eliminated, and plays a vital role in the spin shelving process. The data on dot A point to a subtle back action of the nuclear spins on the electron spin. We speculate that the cycle in the ESR detection, $|\downarrow\rangle \rightarrow |\uparrow\rangle \rightarrow |\uparrow\downarrow,\Uparrow\rangle \rightarrow |\downarrow\rangle$, leads to some alignment of the nuclear field. A significant nuclear polarization is very difficult to achieve with resonant excitation \cite{Imamoglu,Christ} and in fact can be ruled out (a large nuclear polarization would detune the dot from the laser). Instead, it is possible that our experiment aligns the residual Overhauser field, the component resulting from the incomplete cancellation of the fields from each atomic nucleus in the dot, along the $z$-direction. Once the Overhauser field loses its component in the plane, $\gamma_\m{s}$ becomes very small, spin pumping is turned off, and the optical transmission signal reappears. The remarkable feature is the stability of the ESR over the many minutes required to record the data in Fig.\ \ref{esrdotA}. A theory treating the full electron {\em and} nuclear spin dynamics on an equal footing is required \cite{Denon}.

Financial support from the DFG (SFB 631), German Excellence Initiative via ``Nanosystems Initiative Munich" (NIM) and EPSRC (UK) is gratefully acknowledged. We thank J\"{o}rg Kotthaus and Patrick \"{O}hberg for helpful discussions.

\end{document}